\documentclass[epj]{svjour}
\usepackage[T1]{fontenc}
\usepackage[utf8]{inputenc}
\usepackage[english]{babel}
\usepackage{caption}
\usepackage{graphicx}
\usepackage{nicefrac}
\usepackage{amsmath}
\usepackage{amssymb}
\usepackage{amsfonts}
\usepackage{color}

\newcommand{\dime}{0.9\columnwidth}
\newcommand{\ie}{\emph{i.e., }}

\newcommand{\reff}[1]{(\ref{#1})}
\newcommand{\eref}[1]{Eq.\reff{#1}}
\newcommand{\erefs}[1]{Eqs.\reff{#1}}
\newcommand{\figref}[1]{Fig. \ref{#1}}
\newcommand{\figrefs}[1]{Figs. \ref{#1}}

\newcommand{\citet}[1]{\cite{#1}}

\newcommand{\citer}[1]{Ref.\cite{#1}}
\newcommand{\citers}[1]{Refs.\cite{#1}}

\newcommand{\vecb}[1]{\textbf{#1}}

\begin{document}
\title{General features of the linear crystalline morphology of accretion disks}
\author{Giovanni Montani\inst{1,2} \and Brunello Tirozzi \inst{1,2} \and Nakia Carlevaro\inst{1,3}
}                     
%
%
\institute{ENEA, Fusion and Nuclear Safety Department, C.R. Frascati, Via E. Fermi 45, 00044 Frascati (Roma), Italy \and Physics Department, ``Sapienza'' University of Rome, P.le Aldo Moro 5, 00185 Roma, Italy \and Consorzio RFX, Corso Stati Uniti 4, 35127 Padova, Italy}
%
%
\abstract{
In this paper, we analyze the so-called Master Equation of the linear backreaction of a plasma disk in the central object magnetic field, when small scale ripples are considered. This study allows to single out two relevant physical properties of the linear disk backreaction: (i) the appearance of a vertical growth of the magnetic flux perturbations; (ii) the emergence of sequence of magnetic field $O$-points, crucial for the triggering of local plasma instabilities. We first analyze a general Fourier approach to the solution of the addressed linear partial differential problem. This technique allows to show how the vertical gradient of the backreaction is, in general, inverted with respect to the background one. Instead, the fundamental harmonic solution constitutes a specific exception for which the background and the perturbed profiles are both decaying. Then, we study the linear partial differential system from the point of view of a general variable separation method. The obtained profile describes the crystalline behavior of the disk. Using a simple rescaling, the governing equation is reduced to the second order differential Whittaker equation. The zeros of the radial magnetic field are found by using the solution written in terms Kummer functions. The possible implications of the obtained morphology of the disk magnetic profile are then discussed in view of the jet formation.
} 
\maketitle

\section{Introduction}
The Standard Model of accretion in astrophysics \citet{Sh73,SS73} is based on the idea that the angular momentum transport across the plasma disk is ensured by effective viscosity, due to the turbulence phenomena triggered by the magneto-rotational instability (MRI) \citet{BH98}, see also \citers{Ba95,MCP16,CM17}. However, the necessity to account for a magnetic field, due to the central object, together with the long-wavelength response of the plasma disk, implies that the azimuthal component of the generalized Ohm law is consistent only postulating an anomalous plasma resistivity, see \citers{BKL01,MC12}. This effective resistivity makes the disk magnetic field limited in amplitude, because of its diffusive nature \citet{BKL01}. As a result, the possibility to deal with the high magnetic field strength required for the jet generation leads to investigate specific scenarios \citet{Sp18}. 

In \citers{Co05,CR06} (see also \citer{MB11pre}), it has been investigated the possibility for a small scale backreaction of the plasma disk to the magnetic field of the central object, outlining the emergence of a crystalline morphology of the magnetic field microscales, \ie a local radial oscillation of the magnetic flux function. In particular, in the limit of a linear backreaction, in \citer{Co05} it has been written down a Master Equation for the magnetic flux function, which is the starting point of the present investigation. The relevance of this new paradigm relies on the possibility to use the $O$-points of the resulting configuration as the sites where the jet formation could start, as discussed in \citers{MC10,TMC13}. 

Here, we develop a detailed general analysis of the linear Master Equation for the crystalline profile of the plasma disk, outlining two main significant features: \emph{i)} performing a general Fourier analysis, we clarify how the decreasing behavior of the plasma backreaction with the vertical quote (discussed in \citers{Co05,CR06}, see also \citer{BMP11}) is actually a peculiar property of the fundamental harmonics in the radial oscillation; \emph{ii)} by analyzing the Master Equation in the framework of a separable solution, we clarify the existence of regions of the parameter space, where a relevant sequence of $O$-points of the magnetic configuration can take place.

Both the results mentioned above have a precise physical implication. The point \emph{i)} shows how the higher order harmonics of the radial oscillation of the flux function are actually associated to an inverse behavior of the vertical gradient between the background and the linear perturbation. This issue calls attention for its possible implications on the morphology of the nonlinear backreaction which is associated with the fragmentation of the disk in a ring-like morphology, see \citer{CR06} and also \citers{MP14aip,MP14pp,MRC18}. The point \emph{ii)} demonstrates that the topology of the induced radial component of the magnetic field can be characterized by a sequence of $O$-points, where such a radial component vanishes (noting that the dipole-like magnetic field of the central object is essentially vertical on the equatorial plane). This feature, absent in the analysis in \citers{Co05,CR06}, suggests that the magnetic micro-structure, emerging when the perturbation scale is sufficiently small (\ie for sufficiently high values of the plasma $\beta$ parameter), could be an interesting mechanism in triggering sites for the jet formation. It is worth noting that we also outline the existence of a parameter region for the model in which both the radial and the vertical dependence of the backreaction magnetic surfaces oscillate.

\section{Standard model for accretion}
The basic idea for accretion onto a 
compact object has been formulated in \citer{Sh73} and \citer{SS73}, where essentially thin disk configurations were considered. 
Actually, thin accretion disks are commonly present around stellar and black hole systems and they have the advantage to elucidate the accretion features in a one-dimensional model, \ie essentially the radial dependence on the $r$ coordinate is described, being the vertical one ($z$) averaged out of the problem and the azimuthal dependence on $\phi$ is forbidden by the axial symmetry of the steady disk configuration \citet{BKL01}.

For such a steady axially symmetric one dimensional representation of a thin disk configuration, accreting material 
onto a compact object, having a mass 
$M$ and an almost vertical magnetic field $\textbf{B} = B\hat{\textbf{e}}_z$, 
is determined via the implementation 
of the mass conservation equation and of 
the momentum conservation ones.

If we define the disk accretion rate as 
\begin{equation}
\dot{M} \equiv - 2\pi r\Sigma u_r \; , 
\label{jp1}
\end{equation}
where $\Sigma$ and $u_r$ denote the 
superficial density and the radial infalling velocity of the fluid, respectively, then the mass conservation equation reduces to the following condition
\begin{equation}
d\dot{M}/dr = 0\; . 
\label{jp2}
\end{equation}

The radial component of the momentum conservation equation splits into 
two components, one stating that the disk angular velocity $\omega$ is 
essentially equal to the Keplerian one 
$\omega_K$ and the advection and the pressure gradient  balance each other separately, \ie 
\begin{eqnarray}
\omega (r) \simeq \omega _K 
= \sqrt{GM/r^3}\;,\label{jp3a}\\
\rho_0 u_r\frac{du_r}
{dr} + \frac{dp_0}{dr} = 0 \; , 
\label{jp3b}
\end{eqnarray}
$\rho_0(r)$ and $p_0(r)$ denoting the 
values of the mass density and of 
the pressure, taken on the equatorial plane of the disk. 

The separation of the gravo-static equilibrium into Eqs. (\ref{jp3a}) and (\ref{jp3b}) is a natural assumption when the thinness of the disk is taken into account, see \citer{BKL01}, and it relies on the idea that the advection and pressure gradient terms balance each other, on a different scale with respect to the Keplerian motion of the disk. In \citer{Ogilvie97}, it has been systematically investigated the equilibrium configuration of a disk, demonstrating that, only when the disk is thick enough, the differential angular velocity significantly deviates from the Keplerian value (a configuration very far from the applicability presented in Eqs.(\ref{jp3a}) and (\ref{jp3b}) is recovered in the so-called ADAF \cite{AF13} and see also \citers{PMB2012,Pugliese_2013} for the hydrostatic equilibrium in strong gravitational field).

When, in the next Section, we will analyze the magnetic micro-structures generated at small spatial scales, we will see that a deviation from the Keplerian rotation of the disk can be also induced by the plasma backreaction and in that case, it is compensated by the Lorentz force. However, there, the advection term is not present, but, as it has been discussed in \citer{MC12}, such contributions, at high $\beta$ values of the plasma, live on a macroscopical spatial scale, decoupling from the microscale balance of the Lorentz force with the correction to the centripetal Keplerian force (see also the discussion in Sect. \ref{susecsec}).

The vertical component of the momentum equation determines the decay law of 
the mass density far away the equatorial plane. The details of such decay depend on the particular equation of state we adopt for the fluid, but a 
reliable expression, valid both for the general 
polytropic and isothermal case, reads as 
\begin{equation}
\rho (r,z) \simeq \rho_0 (r) 
\Big( 1 - \frac{z^2}{H^2}\Big) \; , 
\label{kp4}
\end{equation}
where $H(r)$ denotes the half-width of the disk and if it is thin we must have $H/r \ll 1$. The spatial scale of the disk vertical extension is fixed in the vertical gravostatic equilibrium via the ratio between the sound 
velocity to the Keplerian angular velocity, \ie
$H\sim v_s/\omega _K$. 
The sound velocity can be reliably 
estimated via the relation 
$v_s\sim \sqrt{K_BT/m}$, being $T$ the 
disk temperature, $K_B$ the Boltzmann 
constant and $m$ the proton mass. 

Finally, the azimuthal component of the momentum conservation equation, 
which regulates the angular momentum transport across the disk, provides the following expression of the accretion rate:
\begin{equation}
\dot{M} = 6\pi \eta_v H\; , 
\label{jp5}
\end{equation}
here $\eta_v$ denotes the shear viscosity coefficient, associated to the 
friction of the disk layers in differential rotation with $\omega_K$.
This viscosity effect can not be due to the kinetic conditions of the plasma, 
which is actually quasi-ideal (see \citer{MP13}), and in \citer{Sh73} $\eta_v$ has been justified via 
a turbulent behavior emerging in the disk spatial microscales. 
Such a coefficient can be estimated by comparing the exact expression 
of the $(r,\,\phi)$ component of the viscous stress tensor $\tau _{ij}$ with its 
interpretation in terms of the correlation function of the turbulent radial and azimuthal velocity component $v_r$ and $v_{\phi}$, respectively, \ie
we have 
\begin{equation}
\tau _{r\phi} = \eta_vr\frac{d\omega_K}{dr} = - \frac{3}{2}\eta _v\omega _K= - \langle \rho_0 v_rv_\phi\rangle 
= - \alpha \rho _0 v_s^2\; , 
\label{jp6}
\end{equation}
from which we get the basic Shakura 
expression 
\begin{equation}
\eta_v = \frac{2}{3}\alpha \rho _0 v_sH\; . 
\label{jp7}
\end{equation}
The parameter $\alpha \le 1$ has been introduced to phenomenologically account that all the supersonic velocity fluctuations are unavoidably damped in the turbulent regime. 

By combining this expression for $\eta _v$ with the relation \reff{jp5}, 
we arrive to the final expression for the accretion rate of the disk in terms 
of basic quantities (particle density, temperature, central body mass), namely 
\begin{equation}
\dot{M} = 4\pi \alpha \rho_0 v_sH^2 
= 4\pi \alpha \rho _0 \omega _K H^3\;.
\label{jp8}
\end{equation}

The possibility to deal with a turbulent behavior of the plasma requires 
that a sufficiently significant instability can affect the equilibrium configuration. Since in the case of 
a Keplerian disk the steady profile turns out to be stable \citet{BH98}, after some years of investigation, the source of turbulence has been identified in the MRI \citet{HB91,Ba03}. Firstly derived in \citer{Ve59} and \citer{Ch60}, 
MRI is an Alfv\'enic instability taking place when the magnetic field intensity within the plasma is below a 
given threshold. In particular, for 
a Keplerian disk, the simplest condition to 
deal with MRI reads as follows
\begin{equation}
\omega_A \equiv kv_A < \sqrt{3}\;\omega_K\; , 
\label{jp9}
\end{equation}
$v_A$ being the Alfv\'en speed and $k$ 
the perturbation wave-number.
In correspondence to an assigned value of the magnetic field, a minimal 
spatial scale exists $\lambda = 
2\pi /k$ for which MRI holds. 
Such a scale, say $\lambda_{min}$ 
is easily identified as 
\begin{equation}
\lambda_{min} \simeq \frac{2\pi}{\sqrt{3}} 
\frac{v_A}{v_s} H \sim 
\frac{H}{\sqrt{\beta}}\; , 
\label{jp10}
\end{equation} 
$\beta$ denoting the standard plasma parameter, which in astrophysics can take also rather large values. 

In \citer{Co05}, it has been shown that just 
on the small scale $\lambda_{min}$, 
there stated as $\sim v_A/\omega_K$, 
the plasma backreaction to the central body can also take a very different nature with respect turbulence. 
In fact, a steady perturbation to the equilibrium is allowed for which the correction of the centripetal force associated to the differential rotation, is directly linked to the perturbed 
Lorentz force raised in the plasma 
by the emergence of current densities 
on very small scales. We observe that the 
background magnetic field is current-free since it is due to the central body magnetosphere and, within the disk, 
it is a vacuum field. In \citer{CR06}, this idea of a microscopic crystalline morphology of the perturbed magnetic field, \ie its radial oscillating behavior on the scale 
$\lambda_{min}$, is implemented also in the limit of a nonlinear beackreaction. As a result, it is possible to show that the disk morphology is decomposed in a series of microscopic ring-like structures, \ie the mass density 
of the disk acquires periodic nodes, see also the global model developed in \citer{MB11pre}.

In the simplest case of a linear small scale backreaction within the disk, the magnetic flux function is currugated and 
this perturbed profile is associated to 
a linear two-dimensional partial differential equation, dubbed in what follows as Master Equation for the disk crystalline structure.
We will discuss in detail the morphology of the solutions associated to this 
equation in order to extract information about the small scale properties 
of the disk steady profile on which MRI can develop. We conclude by observing that in \citer{MC12}, 
it was argued how the presence of a small scale crystalline disk backreaction and the associated relevant micro-current density can have significant implications on the problem of the ``anomalous'' disk resistivity, required to 
account for the observed accretion rates in systems like X-ray binary stars.

\section{Physical model}\label{sec3}
We consider a steady and axialsymmetric thin disk configuration, 
characterized by a magnetic field $\vecb{B}$ having the following poloidal form
\begin{equation}
\vecb{B} = -r^{-1}\partial_z \psi 
\hat{\textbf{e}}_r + r^{-1}\partial _r\psi 
\hat{\textbf{e}}_z 
\, , 
\label{ti1}
\end{equation}
where we use the aforementioned standard cylindrical 
coordinates $(r\, ,\phi\, ,z)$ 
($\hat{\textbf{e}}_r$, $\hat{\textbf{e}}_{\phi}$ and 
$\hat{\textbf{e}}_z$ being their versors) 
and $\psi (r,z)$ is the magnetic flux function.
The disk also possesses a purely azimuthal velocity field $\vecb{v}$, \ie
\begin{equation}
\vecb{v} = \omega (\psi) r \hat{\textbf{e}}_{\phi}
\, , 
\label{ti2}
\end{equation}
where the angular velocity $\omega$ 
is a function of $\psi$ because we are 
in the range of validity of the so-called 
co-rotation theorem \citet{Fe37}, \ie \erefs{ti1} and \reff{ti2} holds together with the stationary and axial symmetry hypotheses. 

Since, we are assuming the plasma disk 
is quasi-ideal (actually it is true in many range of observed mass density and temperature), 
we neglect the poloidal velocities, 
especially the radial component, 
which are due to effective dissipation, 
according to the Shakura idea of 
accretion discussed in the previous Section. 
Furthermore, we observe that the co-rotation condition (\ref{ti2}) prevents the 
emergence of an azimuthal magnetic field 
component via the dynamo effect.

We now split the magnetic flux function 
around a fiduciary radius $r = r_0$, 
as follows
\begin{equation}
\psi = \psi _0(r_0) + 
\psi _1 (r_0, r - r_0, z) 
\, , 
\label{ti3}
\end{equation}
where $\psi_0$ is the vacuum contribution of the central object around which the 
disk develops (essentially a vertical 
magnetic field comes out from the 
dipole-like nature of the field and 
from the thinness of the disk), 
while $\psi_1$ is a small (still steady) 
correction, here considered of very small 
scale with respect to the background quantities.
By other words, we are studying a small backreaction of the plasma which is embedded in the central object magnetic field, 
whose spatial (radial and vertical) 
scales are sufficiently small to 
produce non-negligible local currents 
in the disk. 

According to \erefs{ti3} and \reff{ti1}, also the 
magnetic field is expressed as 
$\vecb{B} = \vecb{B}_0 + \vecb{B}_1$. 
According to the validity of the co-rotation theorem, at 
any order of perturbation of the steady configuration, 
we expand the angular velocity as follows
\begin{equation} 
\omega (\psi) \simeq \omega _0(\psi _0) +   
\left(d\omega/d\psi\right)_{\psi = \psi _0}\psi _1
\, . 
\label{tiri5}
\end{equation}

In \citer{Co05} (see also \citers{CR06,MB11pre}), it was shown that, 
in the linear regime, \ie $|\vecb{B}_1|\ll|\vecb{B}_0|$, the equilibrium configuration near $r_0$ reduces to the radial equilibrium only, which, at the 
zeroth and first order in $\psi$, gives the following 
two equations 
\begin{align} 
&\omega _0(\psi_0) = \omega _K \, , 
\label{ti6a}\\ 
&\partial ^2_r \psi _1 + \partial^2_z\psi _1 = -k_0^2( 1 - z^2/H^2)\psi _1\, , 
\label{ti6b}
\end{align}
respectively. We recall that $H$ denotes the half-depth of the disk and 
$k_0\sim1/\lambda_{min}$ is the typical wave-number of the small scale
backreaction, taking the explicit form
\begin{equation} 
k_0^2 \equiv 3\omega_K^2/v_A^2 
\, , 
\label{tiri7}
\end{equation} 
and, for a thin isothermal disk, it results 
$k_0H = \sqrt{3}\beta \equiv 1/\epsilon$. As postulated above, in order to deal with small scale perturbations, we have to require that the value of 
$\beta$ is sufficiently large. 

In order to study the solutions of \eref{ti6b}, it is convenient to introduce dimensionless 
quantities, as follows
\begin{equation} 
	Y \equiv \frac{k_0\psi _1}{\partial _{r_0}\psi _0}
\, ,\;\; x \equiv k_0 \left( r - r_0\right) 
\, ,\;\; u = z/\delta\, , 
\label{ti8}
\end{equation}
where $\delta ^2 = H/k_0$. 
Hence, we get
\begin{equation} 
\partial ^2_xY + \epsilon \partial ^2_uY = -\left( 1 - \epsilon u^2\right) Y 
\, , 
\label{t}
\end{equation} 
which, in what follows we dub the Master Equation for 
the crystalline structure of the plasma disk. We recall that, according to \eref{kp4}, the $u$ variable takes a finite range across the thin disk configuration. Moreover, it is also easy to check the validity of the 
relations
\begin{align} 
&B_z = B_{0z}\left( 1 + \partial _xY\right) 
\, , 
\label{ti9a}\\
&B_r\equiv B_{1r} = -B_{0z} \sqrt{\epsilon}\partial _uY
\, , 
\label{ti9b}
\end{align} 
where the existence of linear perturbation regime 
requires $|Y|\ll1$.
 
It is relevant to investigate the solutions of 
\eref{t} in view of determining the
physical conditions under which the crystalline 
structure, discussed in \citers{Co05,LM10}, can actually take place.

\subsection{Reconciling accretion with magnetic micro-structures}\label{susecsec}

Let us now clarify how the magnetic micro-structures can be relevant for the accretion picture of plasma disk. Actually, in \citers{Co05,CR06,MB11pre} the profile of the backreaction on small scale has been analyzed in the presence of differential angular velocity only, i.e. without any poloidal matter flux and therefore accretion features are not addressed. 

In the original idea, proposed in \citers{CR_EPS07,MB09_MGPARIGI}, the formation of magnetic micro-structures was intended to be an intriguing paradigm in which accretion could be sustained also in absence of dissipative effects, only on the basis of the ideal plasma morphology. The crucial point of this reformulation consists of the possibility that, in the case of nonlinear plasma backreaction, a large number of X-points can form. In fact, when the vertical magnetic field is dominated by the backreaction, it structure is intrinsically characterized by a radial oscillation and therefore X-points appear with the same periodicity of the magnetic surface oscillation. Clearly, near such points, the plasma disk manifests its porosity, simply because the prescription of the ideal electron force balance $v_rB_z = 0$ allow now for non-zero radial velocity also in the absence of turbulent viscosity. Of course, in order such a plasma porosity becomes an efficient tool for the disk accretion, a mechanism able to pump the plasma into the X-point is required. In \citer{CR_EPS07}, it was postulated that the pumping of plasma could be supported by intermittent ballooning instabilities, i.e. modulating a naturally mechanism observed in laboratory plasma physics to the astrophysical context scenario. 

Furthermore, in \citer{MC12}, it was emphasized how, limiting attention to a one-dimensional model, a natural inconsistency appears between the Shakura idea of accretion and the emergence of magnetic micro-structures. The nature of this incoherent formulation of the small scale plasma backreaction into a visco-resistive accretion scenario can be easily recognized by observing that, in the Ohm law, the largest contribution to the radial infalling velocity comes from the induced oscillating current densities and therefore, it has, in turn, an oscillating character too (which contradicts the smooth Shakura radial infalling). In a more recent two-dimensional formulation \cite{MC20} (see also \citer{MB11grg}), it has been reconciled the disk magnetic micro-structure with the Shakura scenario of accretion, by making use of the smallness of the poloidal velocity field with respect to the toroidal differential rotation of the disk. Addressing the generalized Grad-Shafranov equation for an accretion disk (see \citer{Ogilvie97}), we have overcome the inconsistency observing that the contribution of the microscale phenomenon must be averaged out on the disk macroscales. As a result, the effect of the current induced by the backreaction allows a proper balance of the Ohm law, even in the absence of resistivity high values, but the induced poloidal velocity is averaged to zero on a macroscopical scale. 

This scenario suggests that the emergence of magnetic micro-structures across the accreting disk could provide a viable alternative paradigm to the necessity of the so-called disk anomalous resistivity, without rejecting the solid scenario of angular momentum transport as driven by the effective viscosity associated to the turbulence that MRI is able to trigger. However, the presence of the small scale backreaction requires very high values of the plasma $\beta$ parameter and therefore there is a disk temperature interval in which MRI can still survive, but the disk is too cold to manifest magnetic micro-structures. In such a region of the disk parameters, the necessity of an effective large resistivity to account for accretion onto compact objects, like X-rays binary stars, can probably no way avoided.

\section{A viable solution to the Master Equation}
We investigate the solution of \eref{t} for the crystalline morphology
where we consider the regime $\epsilon\ll 1$, 
corresponding to large values of the $\beta$ plasma 
parameter. Since the crystalline structure is 
associated  to a periodicity in the 
$x$-dependence, we naturally search 
a solution for the dimensionless first-order magnetic flux function $Y$ in the form
\begin{equation}
Y(x,u) = \sum _{n=1}^{n=\infty} 
F_n(u) \sin (nx + \varphi_n(u)) 
\, , 
\label{t2}
\end{equation}
where $F_n$ and $\varphi_n$ are amplitude and phase, respectively. Above, we included a $u$-dependent phase $\phi_n(u)$ in order to properly weight both $\sin$ and $\cos$ functions in the Fourier expansion. Once set to zero the coefficients of the different trigonometric functions, this expression provides the two following equations
\begin{align}
&2\frac{dF_n}{du}\frac{d\varphi _n}{du} 
+ F_n\frac{d^2\varphi _n}{du^2} 
= 0\, , \label{t3a} \\
&\epsilon
\frac{d^2F_n}{du^2} 
-\epsilon F_n \Big(\frac{d\varphi _n}{du}\Big) ^2 = 
-\left[ (1-n^2) -  \epsilon u^2\right] F_n\, .
\label{t3b}
\end{align}
\eref{t3a} admits the solution
\begin{equation}
d\varphi _n/du = \Theta/F_n^2\,,
\label{t44}
\end{equation}
with $\Theta=const.$, which reduces \eref{t3b} to the following closed form in $F_n$
\begin{equation}
\epsilon
\frac{d^2F_n}{du^2} 
-\epsilon \frac{\Theta^2}{F_n^3} = 
-\left[ (1-n^2) - \epsilon u^2\right] F_n\, . 
\label{t4}
\end{equation}

We observe that, if the periodicity 
is not associated to the fundamental wave-number $k_0$ but to a close one 
$k = \chi k_0$ (where $\chi\simeq1$), the equation above 
is simply mapped by replacing the integer $n$ 
by $\chi n$. For $\epsilon \ll 1$, it is clear that 
we must have $n = 1$ and then we assume 
$1-\chi^2n^2 = \epsilon$. By this choice, \eref{t4} becomes 
independent of $\epsilon$, reading as (noting $F\equiv F_1$)
\begin{equation}
\frac{d^2F}{du^2} - \frac{\Theta^2}{F^3} 
= - \left( 1 - u^2\right) F\, . 
\label{t5}
\end{equation}

We underline how we set $1-\chi^2 n^2 = \epsilon$, thus removing a free parameter from the model, in order to properly recover, for $\Theta = 0$, the same behavior discussed in \citer{Co05}, and it corresponds, from a physical point of view, to deal with a radial wave-number near $k_0$. In the next Section, we will study the Master Equation by using the separation variable method and limiting therefore attention to linear behaviors only. The present case is recovered when the separation constant $\gamma$ obeys the relation $\gamma = 1 - \epsilon$, again ensuring that the radial wave-number, fixed by the $\gamma$ value, remains close to $k_0$. There we will provide a systematic analysis of the mathematical problem for different positive values of $\gamma$ in order to span all the parameter space, without focusing attention to physical constraints. However, it remains clear that, for sufficiently low values of $\gamma$, the short wave-length approximation could break down and, on the contrary, for too large $\gamma$ values, the backreaction scale would become smaller than the Debye length, invalidating the MHD scheme. Such peculiar situations take place in rather extreme conditions and they are fixed by the details of the disk configuration, so that we will not account for them in the next Section.

\begin{figure}
\centering
\includegraphics[width=\dime]{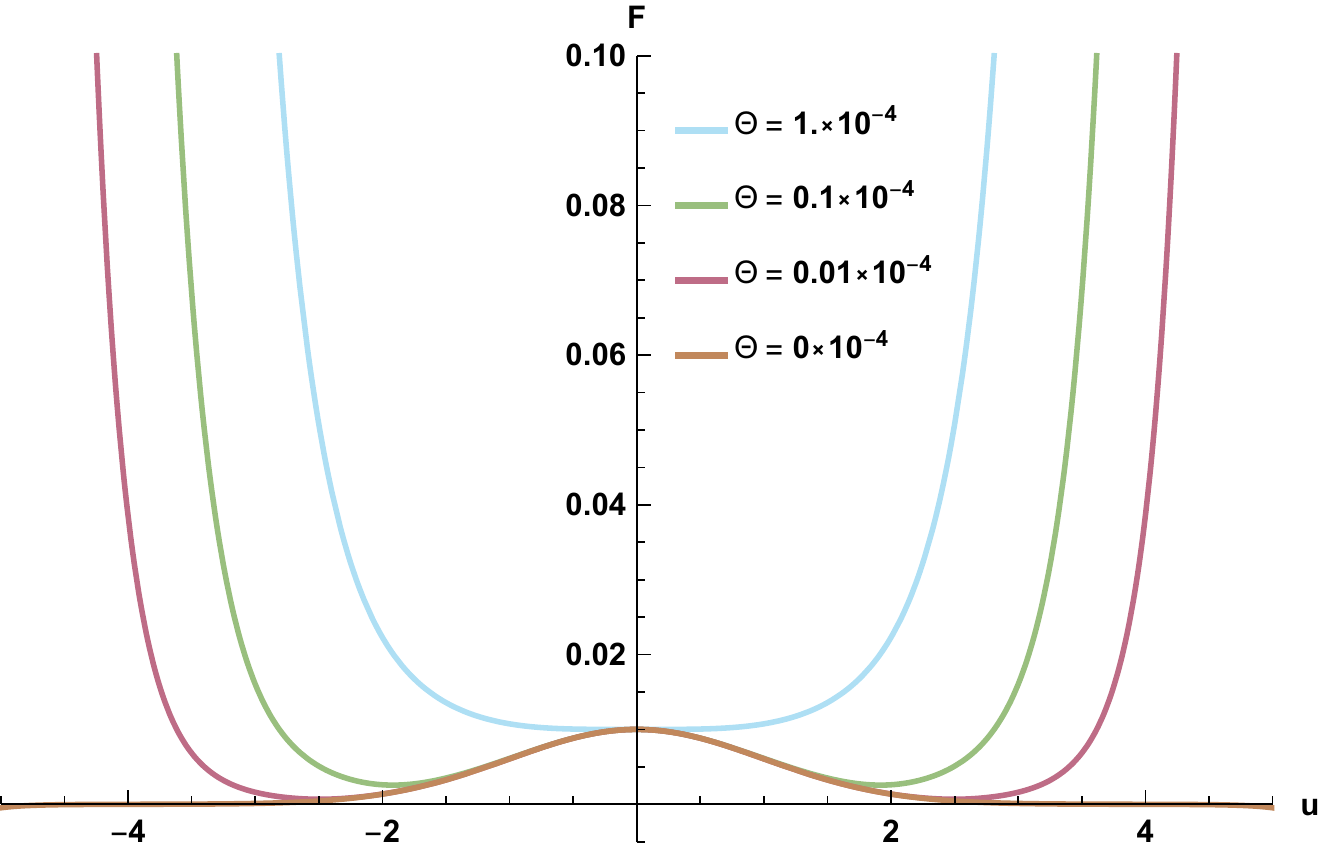}
\caption{Solution of \eref{t5} for the amplitude $F$ of the perturbed flux function as a function of the normalized vertical height of the disk $u$. We fix $\bar{F}=0.01$ varying the parameter $\Theta$ as indicated in the plot.}
\label{1}
\end{figure}

Let us consider the problem of solving \eref{t5} with boundary condition $F(0)=\bar{F}$, being $\bar{F}$ a generic constant (we consider $0<\bar{F}\ll 1$ since such an equation is invariant for $F\to-F$ and because the linear regime must be preserved) and $F'(0)=0$ (here and in the following the prime denotes $u$ derivatives). This last condition has been selected in order to guarantee the symmetry of the solution with respect to the equatorial plane. The plots of the numerical solution, in correspondence of different values of $\Theta$ at fixed $\bar{F}$, are in \figref{1}. Such a solution is a Gaussian for $\Theta=0$ as derived in \citer{Co05}, while for $\Theta\neq0$ it starts to increase for increasing $|u|$ values. It is worth recalling that, when $F$ is greater than unity, the linear approximation fails and the Master Equation is no longer predictive.  

We note that, only for $\Theta = 0$, an exponentially decaying behavior 
of the linear perturbation above the equatorial plane is obtained, 
just like the background profile \citet{BKL01}. Such a natural similarity in the vertical behavior of the background and of the perturbations was also 
at the ground of the analyses in \citers{Co05,CR06,LM10}, but the 
solution derived above demonstrates how 
this feature is a peculiar property of 
the simplest case only, \ie when the 
phases $\varphi_n$ are constant quantities.

We observe that in \eref{t44}, in the linear regime for $F_n\ll 1$, we must require that the constant $\Theta$ be correspondingly small in order to avoid an exceedingly large value of the $u$-derivative of the phase $\phi_n$. Despite the effective $k_z$ value is rather large in the considered scenario (since it is estimated as $k_z \sim \delta^{-1}\sim 1/H\sqrt{\epsilon}$), however, we can explore only region of $u$-values $u\sim k_z$ limited by the request that the linearity of the Master Equation is not broken. In this respect, the increasing behavior of $F$ with large absolute values of $u$ is physically acceptable. Such a behavior can be considered as predictive in the present configurational scheme, only up to the validity of the linearity $Y\ll 1 \Rightarrow F\ll 1$, so that such an increasing behavior is never associated to a diverging non-physical feature: our model is a local one, we are exploring a small region around a fiducial $z_0$ value and only in that neighborhood, our linear equilibrium makes sense. Similar considerations hold also for the radial dependence, when the oscillating behavior were mapped into an increasing one with the radial coordinate $x$, namely the trigonometric function could become hyperbolic ones (see the next Section). 

We also  that, from a physical point of view, the increasing behavior 
is not surprising. In fact, it can be explained because, while the background profile is fixed by a gravostatic equilibrium, the 
perturbation morphology has nothing to do 
with gravity and it is 
mainly produced by the balance of the 
centripetal (say centrifugal) and the 
Lorentz forces.

However, it is important to stress that for sufficiently small values of $\Theta$ a peak of the function $F$ around the equatorial plane is still present. This feature takes place because for small values of $\Theta$ the effect of the nonlinear term $\Theta/F^3$ is minimized. 

From a physical point of view, a process able to 
induce a crystalline structure in the plasma disk 
(for instance a sound or a gravitational wave), \ie a boundary condition able to select the solution associated 
to a radial corrugation of the magnetic flux function, 
should be vertically coherent in phase to produce 
the same vertical gradient of the background profile 
in the perturbation too. On the contrary, a vertical 
shear also in the phase of the underlying process is responsible for an 
inversion of the vertical gradient profile between the 
background and the perturbation. 

The most important physical conclusion of the 
present analysis is to demonstrate that the 
crystalline structure is a very general feature of the 
linearized perturbed equilibrium, but, being associated to 
small values of $\epsilon$ (\ie large values of the plasma 
$\beta$ parameter), the radial oscillation can be associated to a basic wave-number only, very close to the natural one $k_0$.
\begin{figure}
\centering
\includegraphics[width=\dime]{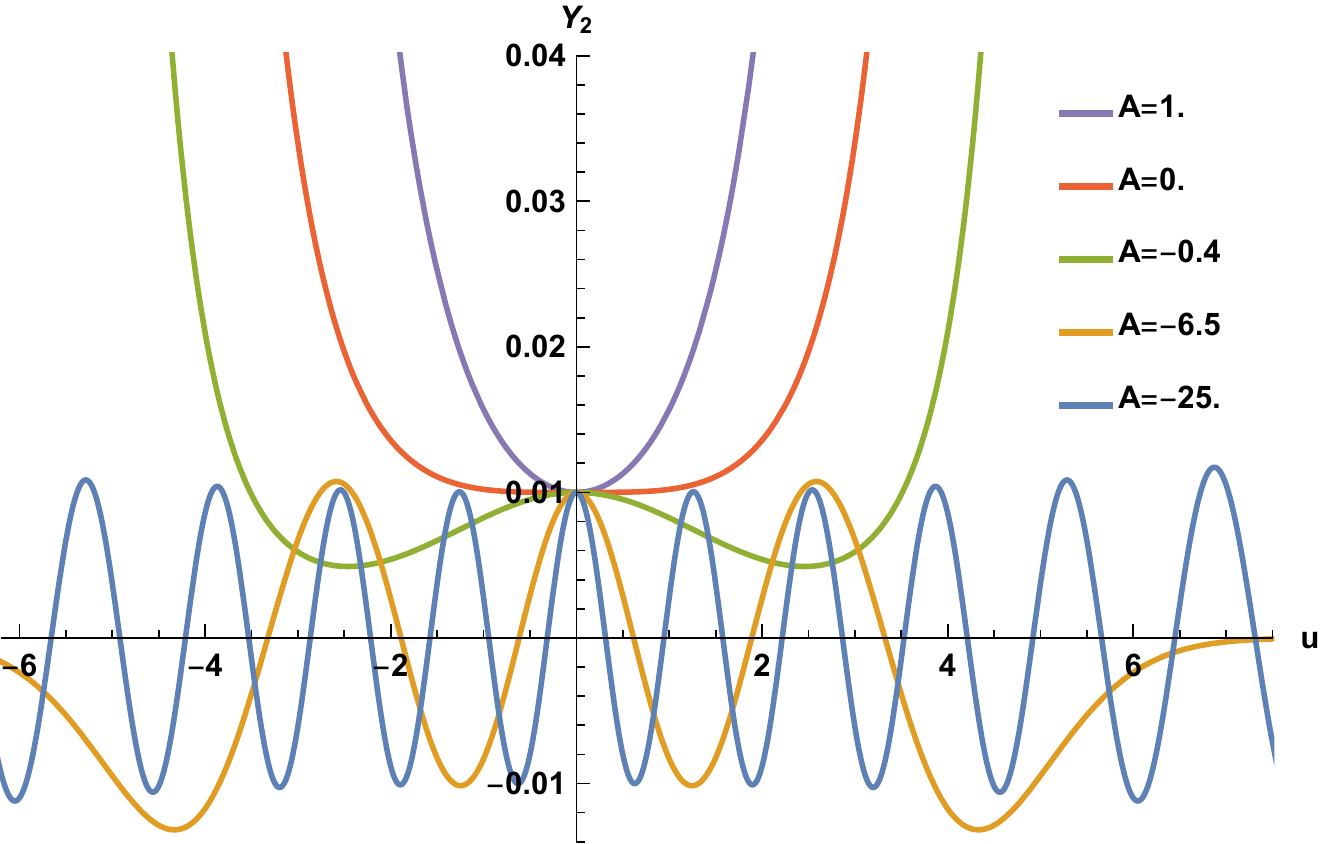}
\caption{Solution of \eref{I3} for the vertical ($u$) dependent part $Y_2$ of the perturbed flux function. We use $Y_2(0)=0.01$ and $Y_2'(0)=0$, by varying the parameter $A$ in $[1,\,0,\,-0.4,\,-6.5,\,-25]$ as indicated in the plot.}
\label{2}
\end{figure}
\begin{figure}
\centering
\includegraphics[width=\dime]{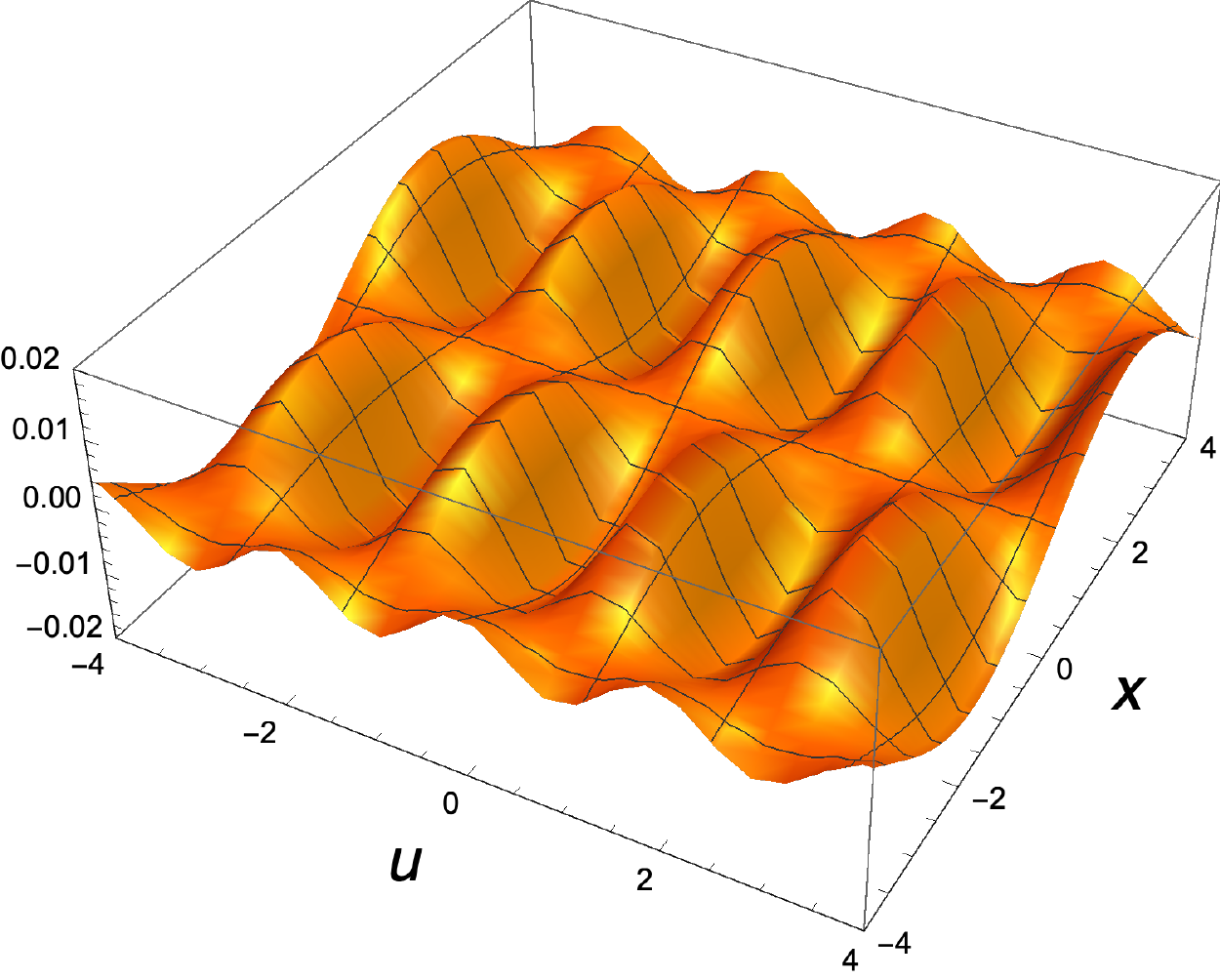}
\caption{Plot of the perturbed magnetic flux function $Y(x,u)=Y_1(x)Y_2(u)$ solution of \eref{hhhgg} (with $C_1=1$ and $D_1=0$) and \eref{I3} for $Y_2(0)=0.01$, $Y_2'(0)=0$, $A=-11$ (\ie $\gamma=0.78$).}
\label{3}
\end{figure}

\section{Separable solution of the Master Equation}
Since the approach in the previous section gave us the solution 
$Y = F(u)\sin(x)$ as the only 
viable case for $\epsilon \ll 1$, 
it is natural to investigate solutions 
relying on the separation of variable method. 
In particular, we will be able to 
explore the situation in which the 
wave-number of the radial oscillation 
($x$-dependence) is a generic constant times 
the fundamental wave-number $k_0$.

In order to investigate solutions relying on the separation of variable method, the Master Equation (\eref{t}) can be rewritten using $Y(x,u)=Y_1(x) Y_2(u)$ in the following decoupled form:
\begin{align}
&(d^2Y_1/dx^2)/Y_1=-\gamma\;,\label{eqy1}\\
&\epsilon(d^2Y_2/du^2)/Y_2+(1-\epsilon u^2)=\gamma\;,\label{eqy2}
\end{align}
where we have introduced the arbitrary constant $\gamma$. The first equation has two possible kind of solutions, namely
\begin{align}
&Y_1(x)= C_1 \sin (\sqrt{|\gamma|} x)+ D_1 \cos (\sqrt{|\gamma|} x),\; \gamma>0\;,\label{hhhgg}\\
&Y_1(x)= C_1 \sinh (\sqrt{|\gamma|} x) + D_1 \cosh (\sqrt{|\gamma|} x) ,\; \gamma<0\;.
\end{align}
\eref{eqy2} can be rewritten in the form of the the Kummer equation of parabolic cylinder functions using the scaling $u\to{u}/\sqrt{2}$. In particular, we get
\begin{align}\label{I3} 
\frac{d^2Y_2}{du^2} - \Big( \frac{1}{4} {u}^2 + A \Big) Y_2=0\;,
\end{align}
where $A=(\gamma-1)/(2\epsilon)$. 
\begin{figure}
\centering
\includegraphics[width=\dime]{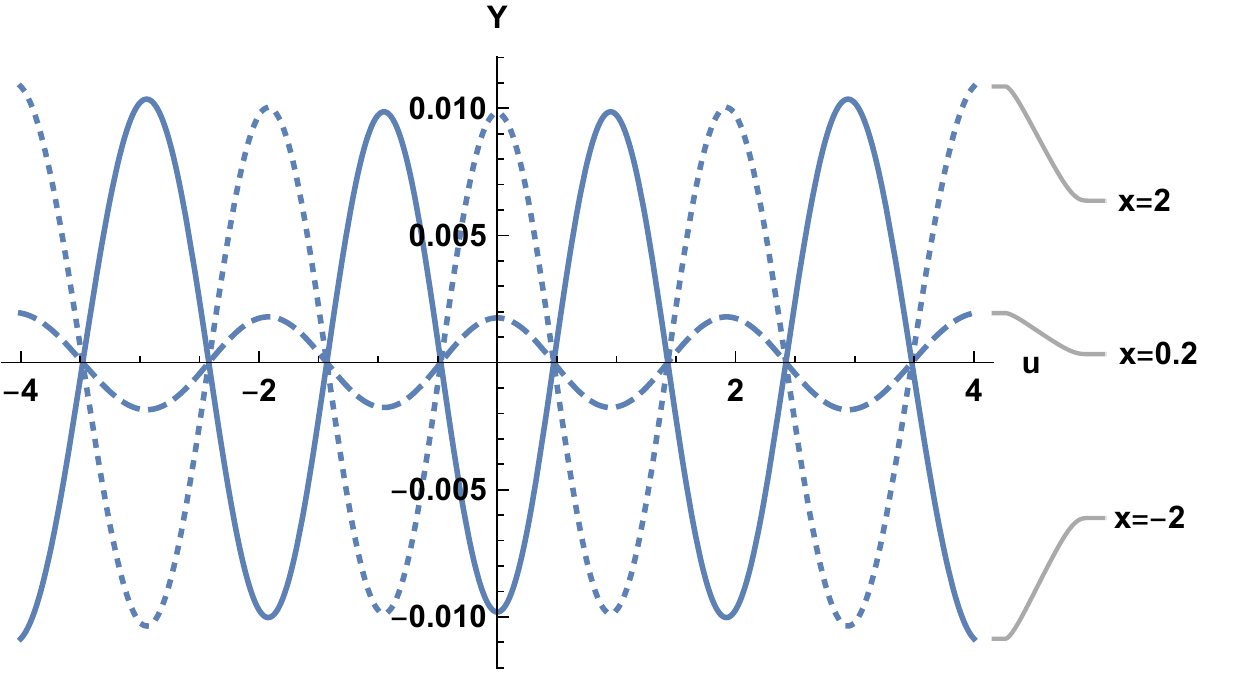}\\
\includegraphics[width=\dime]{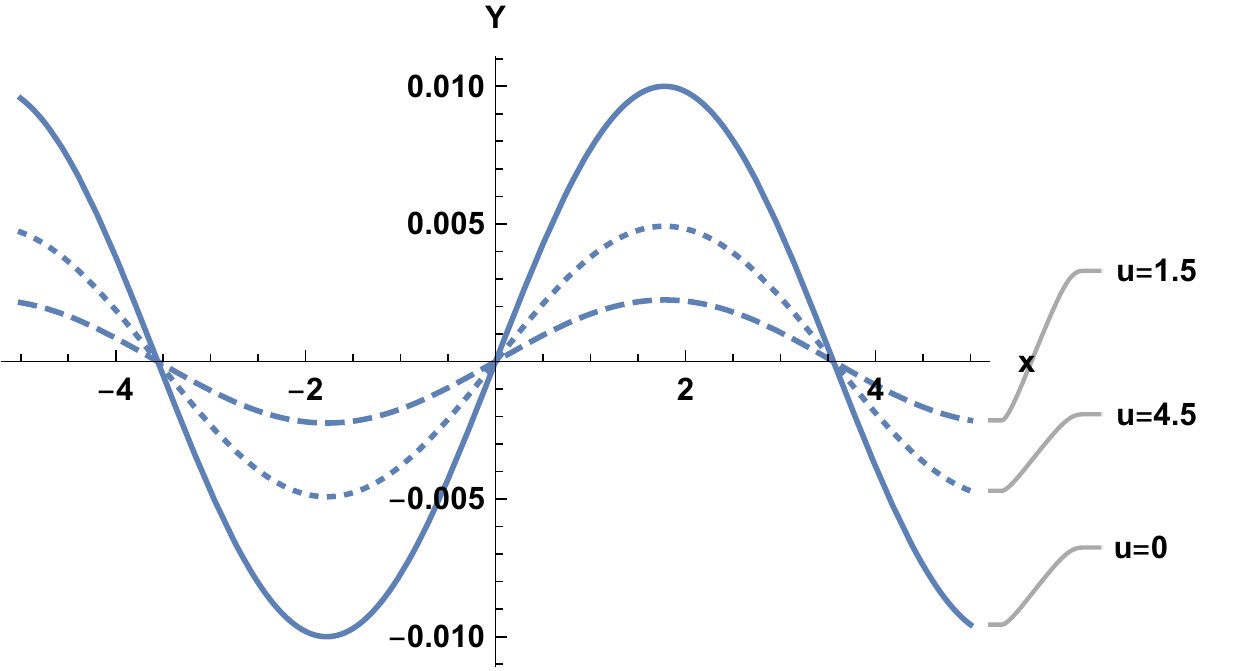}
\caption{Cross sections of the function $Y(x,u)=Y_1(x)Y_2(u)$ form \figref{3} as function of the vertical coordinate $u$ for fixed values of the radial coordinate $x$ (upper panel) and as function of $x$ at given $u$ (lower panel), as indicated in the plots.}
\label{3_1}
\end{figure}
\begin{figure}
\centering
\includegraphics[width=\dime]{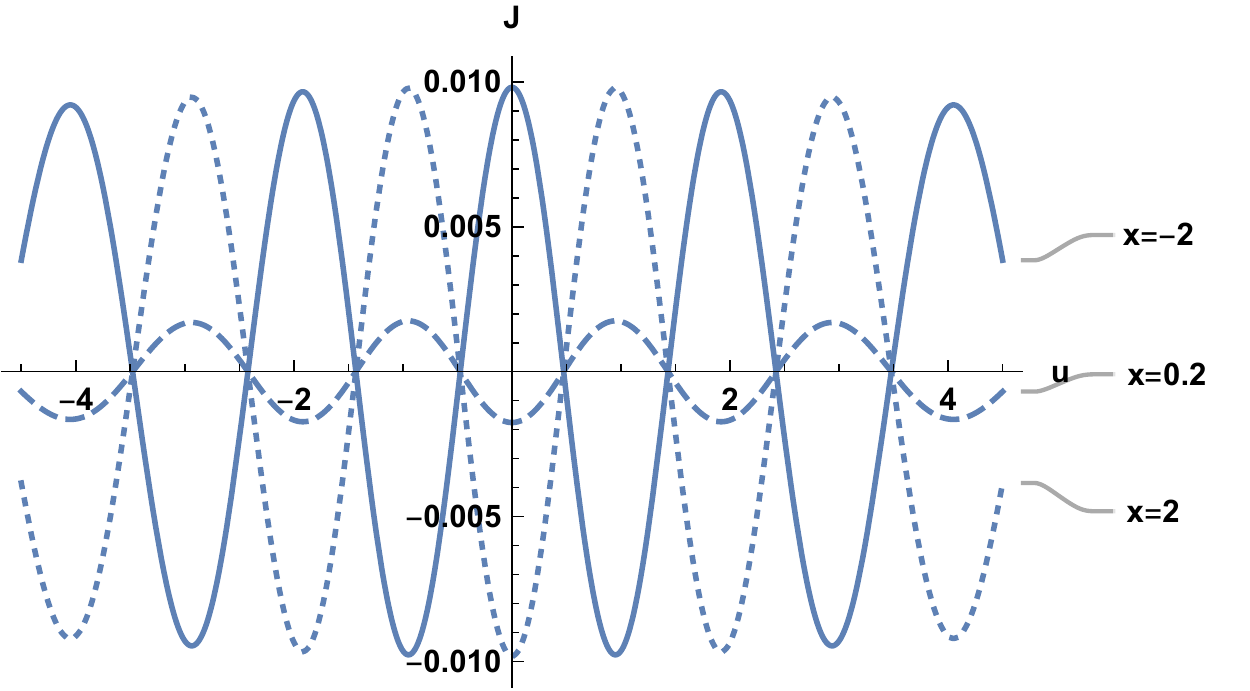}\\
\includegraphics[width=\dime]{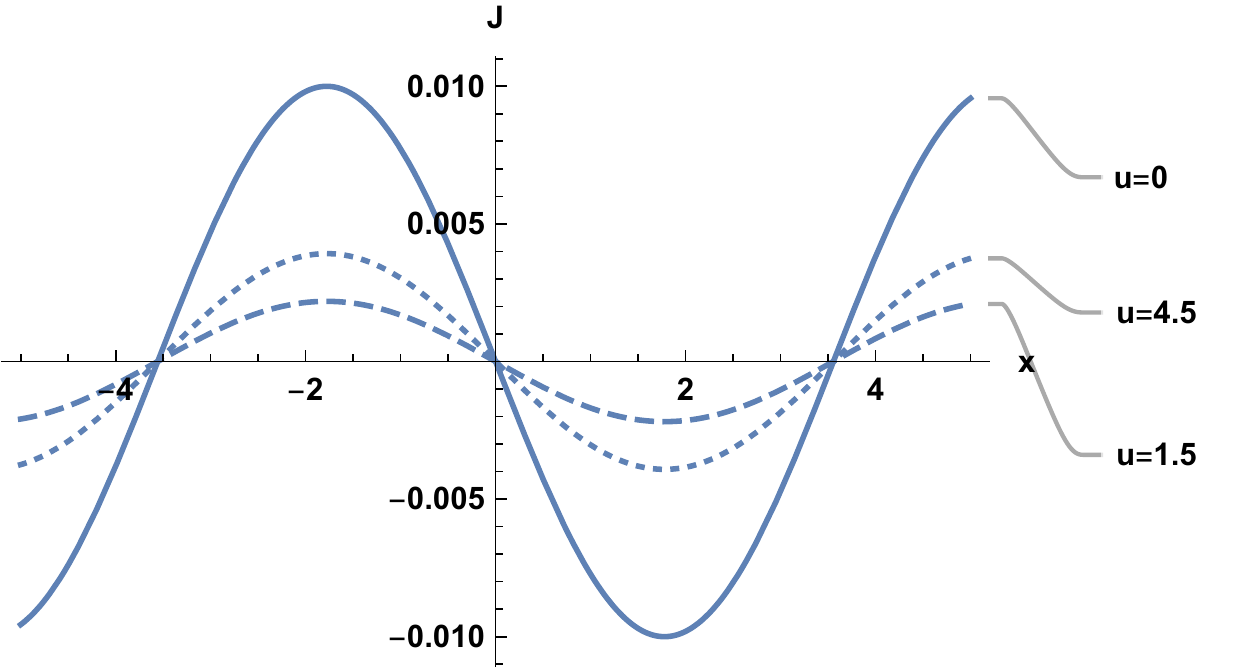}
\caption{Cross section of the current density distribution $J=\partial_x^2 Y+\epsilon\partial_u^2Y$ provided by \eref{t} with the solution of \figref{3}. Panel structure as \figref{3_1}.}
\label{3_2}
\end{figure}

Let us now analyze the properties of the solution by numerically integrating \eref{I3} using initial conditions $Y_2(0)=0.01$  and $Y_2'(0)=0$. We also fix the parameter $\epsilon=0.01$ and run the $\gamma$ parameter in order investigate the morphology of the solution for different values of the constant $A$. 
From the numerical analysis, a range of the parameter $A$ emerges (dependent on the choice of $\epsilon$) for which the behavior of the function $Y_2$ becomes to oscillate, as well sketched in \figref{2}. Since these values of $A$ still corresponds to positive values of 
the constant $\gamma$ (for our setup, $\gamma\lesssim 0.9$), we obtain a range of the solution of the Master Equation where the radial and vertical dependence of the perturbed magnetic flux function both have an oscillating profile as represented in \figref{3} and in the detailed cross sections in \figrefs{3_1} and \ref{3_2}.

Clearly, when $\gamma$ takes negative values, the $x$ dependence 
of the function $Y$ is no longer oscillating (while the $u$-dependence still remains oscillating) and the crystalline 
morphology of the disk is suppressed. As we shall discuss below, the oscillating nature of the vertical dependence of the perturbed magnetic flux function 
has the relevant physical implication that, according to the 
expression for the magnetic field in \erefs{ti9a} and \reff{ti9b}, a series of $O$-points appears in the disk configuration, in which the radial component $B_r$ vanishes.

\section{Whittaker equation and \emph{O}-points}
The many different behaviors of the solution of \eref{I3} can be better understood by investigating peculiar properties of the corresponding analytic solutions. In particular, using the following change of variables: $y={u}^2/2$ and $Y_2=(2y)^{-1/4} W(y)$, \eref{I3} rewrites now as
\begin{equation}\label{W1}
\frac{d^2 W}{d y^2}+\Big(-\frac{1}{4}- \frac{A}{2 y}+\frac{3}{16 y^2}\Big)W=0\;,
\end{equation}
which is a particular case of Whittaker's equation, \ie
\begin{equation}\label{W2}
\frac{d^2 W}{d y^2}+\Big(-\frac{1}{4}+ \frac{K}{y}+\frac{1/4 -\mu^2}{y^2}\Big)W=0\;,
\end{equation}
with $K=-A/2$ and $\mu=\pm1/4$. The linearly independent solutions of this equation are given by
\begin{align}
W^{(1)}_{K,\mu}&= e^{-y/2} y^{\mu+1/2} M(1/2+\mu-K,1+2\mu,y)\;,\\
W^{(2)}_{K,\mu}&= e^{-y/2} y^{\mu+1/2} U(1/2+\mu-K,1+2\mu,y)\;,
\end{align}
where $M$ and $U$ are the Kummer and Tricomi functions \citet{Mi65}, also known as confluent hypergeometric functions of the the first and second kind, respectively. We will use only the function $M$ because the function $U$ is multivalued, and we study the solutions only for finite $y$ since the model holds for a limited range of $u$ values. Moreover, for $\gamma \leq 1$, the function $W_{K,\mu}$ is bounded since $M(1/2+\mu-K,1+2 \mu,y)$ behaves like
$$ \Gamma(1+ 2 \mu) \frac{e^{y/2} y^{-1/2-\mu-K}}{\Gamma(1/2+\mu-K)}+\frac{(-y)^{-1/2-\mu+K}}{\Gamma(1/2+\mu+K)}\;,$$
where $\Gamma$ denotes the standard gamma function. Specifically, the Kummer function $M(a,b,y)$ is formally defined as
$$M(a,b,y)=1+ \frac{a_{(1)}}{b_{(1)}}y+ \dots+ \frac{a_{(n)}}{b_{(n)}}\frac{y^n}{n!}\;,$$
where the rising factorial denoted with the index ${(n)}$ is defined as
\begin{align}
&X_{(0)}=1\;,\nonumber\\
&X_{(n)}= X (X+1)(X+2)\dots(X+n-1)\nonumber\;.
\end{align} 
The sum is converging everywhere, while the derivative of $M$ with respect to $y$ is given by the following recurrent equation: $d M(a,b,y)/dy=(a/b) M(a+1,b+1)$.

%

Using this formalism, we get the following solutions of \eref{I3}:
\begin{align}
\label{jjp}
Y_2^+(u)&=c_1^{+} u \,e^{-u^2/4} M(A/2+3/4, 3/2; u^2/2)\;,\\
Y_2^-(u)&=c_1^{-} e^{-u^2/4}\; M(A/2+1/4,1/2;u^2/2)\;,
\label{jjm}
\end{align}
where $c^{\pm}_{1}$ are integration constants and we denoted with $\nicefrac{+}{-}$ the choice $\mu=\pm1/4$.

The zeros of the derivative of the solutions above correspond to that one of the radial component of the magnetic field from \eref{ti9b}. Using recursive expressions, we get 
\begin{align}
&dY_2^+/du=\nonumber\\
&=c_1^{+} e^{-u^2/4}[(1-u^2/2) M(A/2+3/4,3/2; u^2/2)+\nonumber\\
&+u e^{-u^2/4}(A/3+1/2) u M(A/2+7/4, 5/2; u^2/2)]\;,\label{dy2p}\\
&dY_2^-/du=\nonumber\\
&=c_1^- e^{-u^2/4}(-u/2) M(A/2+1/4,1/2; u^2/2)+\nonumber\\
&+c_1^{-} u e^{-u^2/4}(A+1/2) M(A/2+5/4, 3/2; u^2/2)\;.\label{dy2m}
\end{align}

\begin{figure}
\centering
\includegraphics[width=\dime]{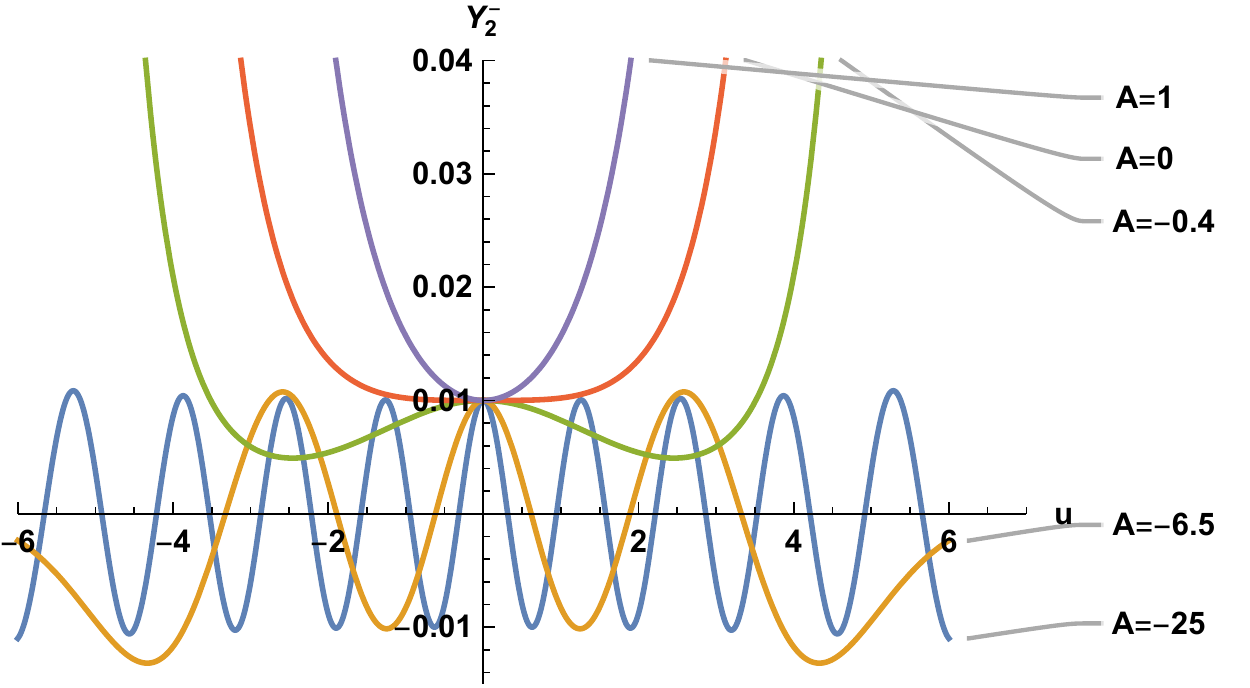}
\caption{Plot of the magnetic flux function component $Y_2^{-}(u)$ in \eref{jjm}, corresponding to $\mu=-1/4$, for different values of the parameter $A$ as in the numerical analysis of \figref{2}. The integration constant $c_1^{-}$ is set to match the boundary condition in $u=0$. }
\label{11}
\end{figure}

\begin{figure}
\centering
\includegraphics[width=\dime]{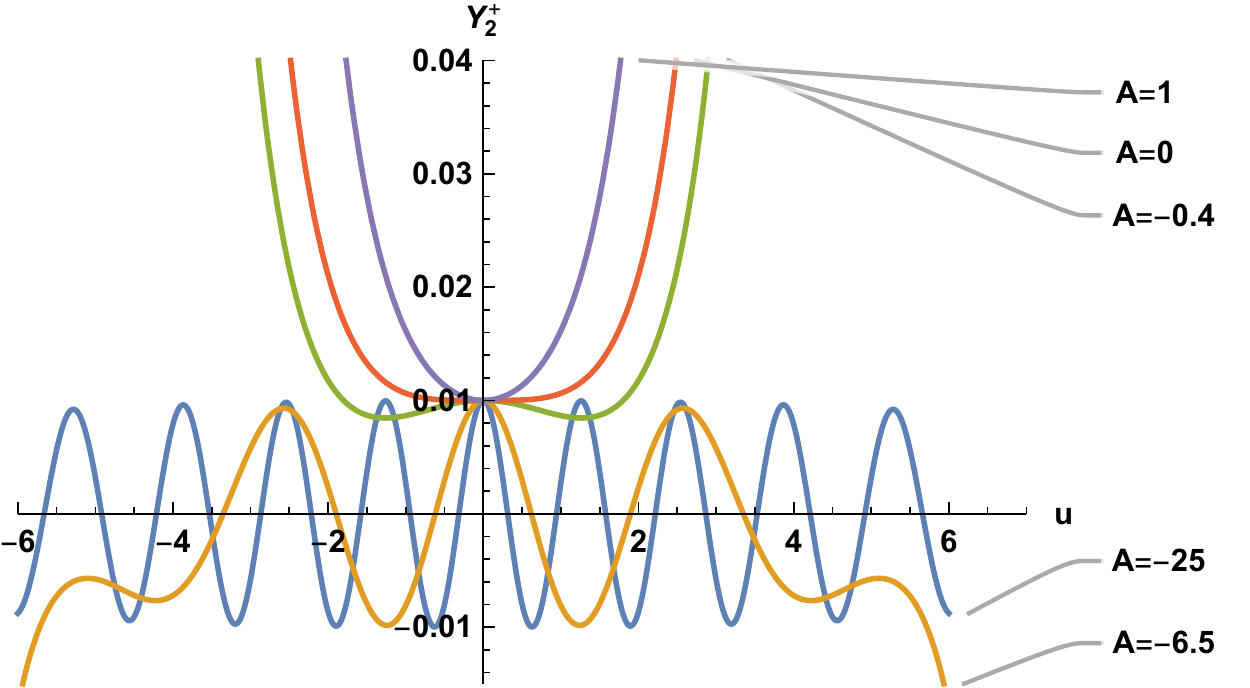}
\caption{Plot of the function $Y_2^{+}(u)$ in \eref{jjp}, corresponding to $\mu=1/4$ (parameter choice as in \figref{11}).}
\label{10}
\end{figure}

\begin{figure}
\centering
\includegraphics[width=\dime]{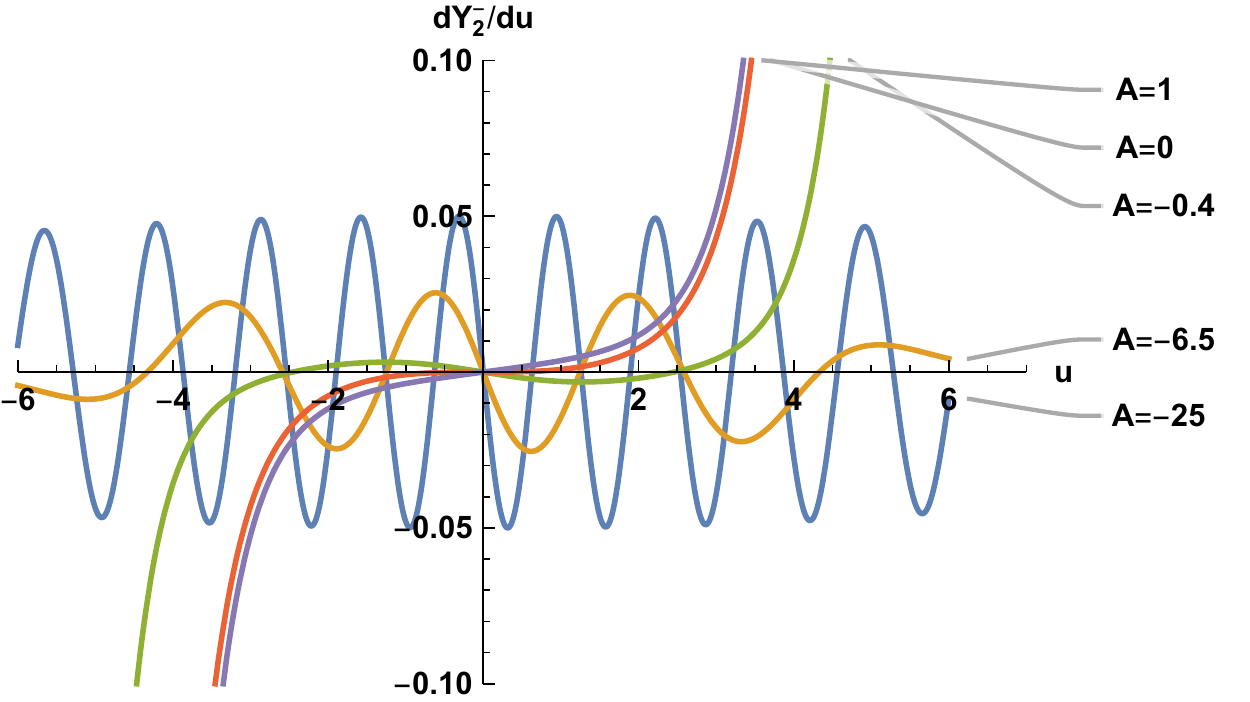}
\caption{Behavior of $dY_2^-/du$ in \eref{dy2m} corresponding to parameter choice of \figref{11}.}
\label{11-}
\end{figure}

Since in the general solution both $\mu=-1/4$ and $\mu=1/4$  have to appear in a linear combination, we plot in \figrefs{11} and \ref{10} the function $Y_2(u)$ for both these values, respectively. Comparing these two pictures with the numerical integration in \figref{2}, we see that the choice $\mu=-1/4$ better addresses the profile associated to the considered physical boundary conditions of the numerical integration. Moreover, it is worth underline the following behavior depending on the values of the $A$ parameter (we focus only on the half-plane $u\geqslant0$): for $-25<A<-16$, $Y_2^-$ oscillates; for $-16<A<0$, the oscillations disappear and the solution has only some maxima or minima; for $A\geqslant0$ it is an increasing function. The transitions of the analytic solution are in agreement with those of the numerical solution of the previous Section. In \figref{11-} we instead plot the graph of $dY_2^-/du$. The $O$-points of the radial component of the magnetic field on the vertical axis clearly emerge, placed in symmetric positions with respect the equatorial plane. For the specific case $A=-25$, the positive zeros of $B_r$ can be found at $u=$[$0$, $0.6$, $1.3$, $1.9$, $2.55$, $3.2$, $3.9$, $...$]. In this respect, we note that algorithms are present for systematically finding the roots of Kummer functions \citet{Mi65}.

%
%
%
%
%
%
%

\subsection{Asymptotic solutions}
Taking into account the solutions for $Y_2(u)$ in \erefs{jjp} and \reff{jjm}, let us now define the following functions
\begin{align}
Z(A,u)=&\Phi_1(u)\cos(\pi(1/4+A/2))+\nonumber\\
&-\Phi_2(u)\sin(\pi(1/4+A/2)\;,\\
V(A,u)=&\frac{1}{\Gamma(1/2-A)}(\Phi_1(u)\sin(\pi(1/4+A/2))+\nonumber\\
&+ \Phi_2(u)\cos(\pi(1/4+a/2))\;,
\end{align}
where 
\begin{align}
&\Phi_1(u)=\frac{\Gamma(1/4-A)}{\sqrt{\pi}\;2^{A/2+1/2}}Y_2^+\;,\\
&\Phi_2(u)=\frac{\Gamma(3/4-A/2)}{\sqrt{\pi}\;2^{A/2-1/2}}Y_2^-\;.
\end{align}
In the asymptotic limit $-A\gg u^2$ (for $A<0$) and finite $u$ (we recall that the thin-disk hypothesis limit the range of $u$ values), we obtain
\begin{align}
&Z(A,u)+i \Gamma(1/2-A) V(A,u)=\nonumber\\
&=\frac{e^{i \pi(1/2+1/2 A)}}{2^{(1/2+1/2 A)}\sqrt{\pi}}\Gamma(1/4- A/2)e^{\theta_r+i \theta_i}e^{i p u}\;,
\end{align}
where $p=\sqrt{-A}$, while $\theta_r(u)$ and $\theta_i(u)$ are assigned polynomials. Thus, the function $Z(A,u)$, the solution $Y_2(u)$ and its derivative with respect to $u$ oscillate giving rise to a sequence of $O$-points across the vertical axis.

For positive values of $A$, we get instead the following asymptotic behavior for $A\gg u^2$:
\begin{equation}
Z(A,u)=\frac{\sqrt{\pi}}{2^{(1/2+1/2 A)}\Gamma(3/4+ A/2)}e^{-\sqrt{A} u+ \theta_1}\;,
\end{equation}
where $\theta_1$ is again an assigned polynomial. This case for $A>0$ is not relevant for the present physical discussion.

\subsection{Laguerre polynomials}
We have shown how, in the case $-A\gg u^2$, there is a countable set of $O$-points, but it is also possible to find other zeros series in the correspondence of smaller $|A|$. It is easy to verify that the following relation takes place between the Kummer function $M(a,b,y)$ and the Laguerre polynomials $L_n^{(\alpha)}$:
\begin{align}
M(-n,\alpha+1,y)=L_n^{(\alpha)}(y)/\binom{n+\alpha}{n}\;.
\end{align}

\begin{figure}
\centering
\includegraphics[width=\dime]{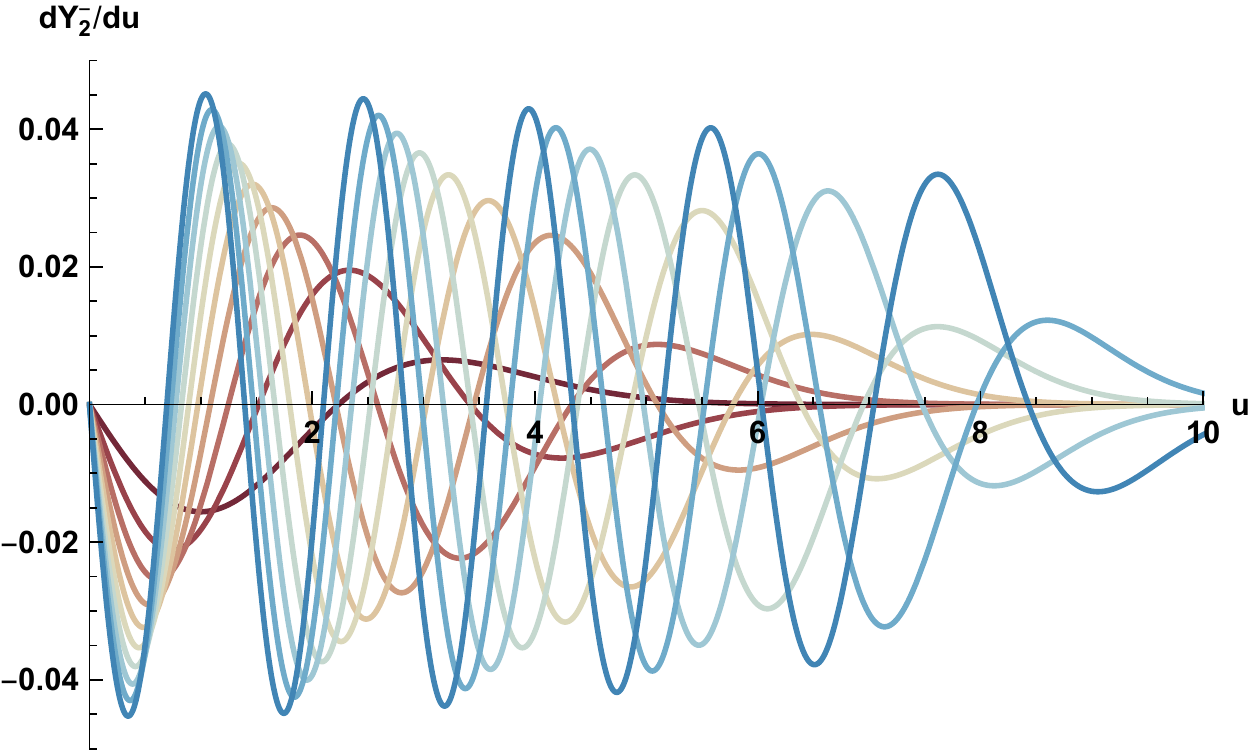}
\caption{Plot of $dY_{2}^-/du$ from \eref{jjq} (zoom of the half-plane $u>0$) by setting integers $n$ in $[1,10]$. Standard color scheme from red ($n=1$) to blue ($n=10$). The number of oscillations increases with increasing $n$.
\label{pape}}
\end{figure}

%
%
%
%
%
%
%
%
%
%
%
For $\mu=-1/4$, \eref{jjm} contains the function $M(A/2+1/4,1/2;u^2/2)$, and we easily get the specific values: $n=-A/2-1/4$ and $\alpha=-1/2$. We thus obtain the following general expression for the derivative of $Y_2^-$:
\begin{align}\label{jjq}
&dY_{2}^-/du=-c_1^- e^{-u^2/4}\times\nonumber\\
&\times\Big(u L_{n-1}^{(1+\alpha)}(u^2/4)+(u/2)L_{n}^{(\alpha)}(u^2/4)\Big)/\binom{n+\alpha}{n}\;.
\end{align}
The equation above is a particular form of \eref{dy2m} and, since it is well-known that for each $n$ there is a finite number of zeros of the Laguerre polynomials which increases with $n$, we have shown how a countable set of zeros on the $z$ axis emerges. In particular, the zeros of Laguerre polynomial $L_n^{(\alpha)}$ belong to the interval $(0, n+\alpha+(n-1) \sqrt{n+\alpha})$, and in \figref{pape} we plot, as an example, $dY_{2}^-/du$ from \eref{jjq} for the first 10 integer $n$ values. 

%
%
%
%
%
%
%
%
%

\section{Discussion}

We have shown how the $z$-dependence of the magnetic 
field is associated to $O$-points of the corresponding configuration. In such points, the radial component of the  magnetic field, due to the plasma backreaction only, 
vanishes. Here, we outline a new feature induced 
by the crystalline morphology of the plasma response 
to the dipole field of the central object.

Let us now discuss the relevance 
of the $O$-points derived above, in view the physical 
properties of the plasma disk thought as an accreting 
structure. Our analysis has been performed for an ideal 
plasma disk in the absence of poloidal 
components of the velocity field, \ie
$v_r = v_z \equiv 0$. However, in the 
Standard Model of accretion \citet{Sh73}, the plasma must have a finite
electric 
conductivity and the azimuthal component 
of the generalized Ohm law reads 
\begin{equation}
v_zB_r - v_rB_z = J_{\phi}/\sigma\;,
\label{aol}
\end{equation}
where $J_{\phi}$ is the corresponding azimuthal 
component of the current density and 
$\sigma$ is the constant coefficient of 
electric conductivity.
In the standard mechanism for accretion 
onto a compact object, both the quantities 
$B_r$ and $v_z$ are essentially negligible 
and, where $\sigma$ takes very large values, the accretion is suppressed, 
\ie since $B_z\neq 0$, we must get 
$v_r\simeq 0$. 

In the spirit of the present analysis, 
in those regions where the plasma is 
quasi-ideal, we find the relation 
\begin{equation}
v_z = v_r B_z/B_r\;. 
\label{old}
\end{equation}
Clearly, nearby the numerous $O$-points 
of the magnetic profile, where $B_r$ is 
nearly vanishing, we get very high values 
of the vertical velocity, which suggests 
the existence of privileged plasma sites for the jet formation. 
This perspective has been investigated in 
some detail in \citers{MC10,TMC13}.

It is worth observing that, when an anomalous resistivity is
considered, the term $v_z B_r$ can be actually safely neglected since the
balance of the Ohm law is guaranteed by the relation $v_rB_z = -
J_{\phi}/\sigma$. Of course, if we can state that $v_zB_r\sim v_rB_z$, we
naturally have large values of $v_z$ (due to the smallness of $B_r$ in a thin disk). However, this situation can not be regarded as the most general one, unless some other information on the velocity field is assigned by extra physics in the disk.
On the contrary, when the values of sigma are very large, the balance
between the two terms is mandatory and $v_z$ is very large near $O$-points.
Clearly, also the solution $v_r\sim v_z = 0$ is viable, but this
situation is just that one discussed in Sect. \ref{sec3}, where differential disk angular velocity is included only. However, in order to get the equality
$v_rB_z = v_z B_r$ and the $O$-points, simultaneously, $J_{\phi}/\sigma$
must be a negligible contribution, \ie
$k_0$ can not exceed a critical magnitude fixed by the value of the electric conductivity $\sigma$.
If not, we can still have a large value of $v_z$, but again in specific
cases only.

Actually, the study of the crystalline micro-structures in the presence of
poloidal velocity has not yet been fully investigated, see \citer{MC20}
for recent developments (the most important difficulty relies on
the nonlinear advection terms in the poloidal velocity components). However, the picture associated to the Master Equation is reliable in view
of the smallness characterizing the poloidal velocity components with respect to the toroidal rotation velocity, as expected in a thin accretion disk.

The main merit of the present analysis 
consists of having outlined the general 
character that the $O$-points takes in the 
linear profile of the perturbed magnetic field, 
as soon as the short wavelength backreaction 
of the plasma is excited in 
disk with high $\beta$ values.

\section{Conclusions}

We analyzed general features of the small scale morphology 
of the backreaction that is generated in a thin plasma, 
embedded in the magnetic field of the gravitationally 
confining central object. In particular, we studied the 
Master Equation of the so-called crystalline structure 
\citet{Co05}, searching for a satisfactory characterization 
of the admissible profile in the linear backreaction limit. 

We have followed two different, but complementary, approaches. The first one is based on a general Fourier expansion, while the second one relies on a separated variable procedure.
This analysis offered a complete spectrum of the 
available solutions, outlining new behaviors in the 
plasma equilibrium, absent in the basic treatment 
in \citers{Co05,CR06}. 

The main merit of this systematic study of the 
linear backreaction of the plasma disk consists 
of the relevant physical implications of the 
obtained solutions, as discussed above, \ie the inverse 
behavior of the vertical gradient with respect to the background one (as the highest harmonics are considered) and the possible 
emergence of a consistent sequence of $O$-points, 
where the radial magnetic field component vanishes. 

The first of these issues is of interest in view of the 
nonlinear scenario of the plasma backreaction discussed in \citers{CR06,MB11pre}, where, see also \citer{LM10}, the possibility for the radial fragmentation of 
the disk into a microscopic array of rings is demonstrated. 
In this respect, the linear inversion of the vertical gradients 
suggests that, in the nonlinear regime, a separation of 
the disk into two symmetric components, up and down the equatorial plane, could take place, with significant implications on the transport features across the disk. 

The second result, concerning the appearance of $O$-point series of the magnetic configuration, is relevant in view of 
the realization of conditions for jet emission. 
In fact, nearby the $O$-points, when the plasma 
resistivity is suppressed, the vertical component of 
the plasma velocity can take very large values, see the discussion in 
\citers{MC10,TMC13} about the seeds of a vertical matter flux from the disk. Using the solution given in terms of hypergeometric confluent functions, we have also analytically shown the existence of the $O$-point series and such a series critically depends on the parameter of the separation of variable technique.

The systematic analysis here pursued of the Master Equation for 
the crystalline structure of a stellar accretion disk, 
constitutes a well-grounded starting point for the 
investigation of nonlinear features of the magnetic 
confinement of the plasma disk. In fact, as discussed in 
\citer{CR06}, the extreme nonlinear regime is characterized by 
a dominant influence of the MHD-force in confining the plasma with respect to the radial gravitational field. Furthermore, in \citer{MBC15} it is outlined how the 
request that the Master Equation holds in the nonlinear case too turns out to be a proper choice to close the equilibrium system, preserving the crystalline morphology.

\section*{Authors contributions}
All the authors were involved in the preparation of the manuscript.
All the authors have read and approved the final manuscript.


\end{document}